\documentclass[showpacs,amsmath,amssymb,prl,aps,twocolumn]{revtex4}
\usepackage{graphicx}
\begin{document}
\title {Spin-Hall effect in a [110] quantum well}
\author {E. M. Hankiewicz and  G. Vignale}
\affiliation{Department of Physics and Astronomy,and University of
Missouri, Columbia, Missouri 65211, USA}
\email{hankiewicze@missouri.edu}
\author{M. E. Flatt\'e}
\affiliation{Department of Physics and Astronomy, University of
Iowa, Iowa-City, IA 52242}
\date{\today}
\begin{abstract}
A self-consistent treatment of the spin-Hall effect requires
consideration of the spin-orbit coupling and electron-impurity
scattering on equal footing. This is done here for the
experimentally relevant case of a [110] GaAs quantum well [Sih
{\it et al.}, Nature Physics 1, 31 (2005)].   Working within the
framework of the exact linear response formalism  we calculate the
spin-Hall conductivity including  the Dresselhaus linear and cubic
terms in the band structure, as well as the electron-impurity
scattering and electron-electron interaction to all orders. We
show that the spin-Hall conductivity naturally separates into two
contributions, skew-scattering and side-jump,  and we propose an
experiment to distinguish between them.
\end{abstract}
 \maketitle

{\it Introduction.}  The current focus of spintronics based on
spin-orbit (SO) interactions is on the spin-Hall effect
\cite{Dyakonov71,Hirsch99,Murakami03,Sinova04} (SHE), i.e. the
generation of a steady {\it spin current} transverse to a d.c.
electric field.   The recent experimental observations of the SHE
\cite{Kato04,Wunderlich05,Sih05} have enhanced the interest of the
community in this topic. At first sight the SHE seems to arise
from two quite different mechanisms: one associated with SO
interactions between electrons and impurities
\cite{Dyakonov71,Hirsch99,Zhang00,Hankiewicz05}, and the other
connected with SO interactions in the band structure of the
material~\cite{Murakami03,Sinova04}. In reality, the two
mechanisms are inseparable. On one hand, the presence of
electron-impurity scattering is essential to ensure that the
system reaches a steady state in the presence of the electric
field. This means that the d.c limit ($\omega \to 0$,
 where $\omega$ is the
frequency of the electric field) must be taken before letting the
electron-impurity scattering time $\tau$ tend to infinity in the
``clean limit".  On the other hand,  the unitary transformation
that reduces the original multi-band Hamiltonian of the solid to
an effective Hamiltonian for, say, the conduction band,   not only
generates SO coupling terms (for example the Dresselhaus term) but
also  modifies the position operator (see Eq.~(\ref{r.effective})
below). All the above effects have an impact on the spin Hall
conductivity (SHC). The so-called skew-scattering (SS)
contribution\cite{Smit55} arises from the asymmetry of the
electron-impurity scattering in the presence of SO
interactions\cite{Mott}.  The   ``side-jump"
(SJ)\cite{Berger70a,Lyo72,Nozieres} comes from the change in the
form of the position and velocity operators.    Finally, the
``intrinsic" ~\cite{Murakami03,Sinova04} contribution arises from
the SO coupling terms in the effective Hamiltonian. Notice that
the side-jump and the intrinsic contributions have a common origin
in the unitary transformation that reduces the multi-band
Hamiltonian to an effective one-band Hamiltonian.

Although all types of contributions are present in experiments
\cite{Kato04,Wunderlich05,Sih05}, theoretical approaches usually
focus on one or the other.  Some papers focus on the  band
structure effects neglecting the effect of spin-orbit coupling on
electron-impurity scattering \cite{Inoue04,Halperin04}; others on
the SS and SJ effects, ignoring spin-orbit interactions in the
band structure \cite{Engel05,Sarma06,Hankiewicz05}. A complete
theory should of course describe all contributions on equal
footing~\cite{Sarma06b}.  Furthermore, it would be desirable to
 have a way to distinguish experimentally the various contributions.

In this paper we present a complete theory of the SHE for an
experimentally well-studied system -- a GaAs [110] quantum well
(QW)\cite{Sih05} -- in which the electron-impurity and the
electron-electron interaction can be studied simultaneously and
consistently with the spin-orbit terms in the band structure,
namely the  linear and cubic Dresselhaus terms. Let us emphasize
that in this system the spin-current and the SHC are well defined
 because the $z$-component of the spin is conserved. We show
 that the Dresselhaus terms do not contribute
 while  the  conservation of $S_z$ component implies  that the
SHC includes only ``pure" skew scattering and side jump
contributions. Furthermore, we are able to prove (using exact
linear response theory) that the side-jump contribution is
independent of the strength of disorder and Coulomb interactions
-- a fact that had been shown on the basis of perturbative
calculations to first order in disorder
~\cite{Luttinger,Nozieres,Bruno01},
 but never before shown to be true at all orders in the strength of the interactions and disorder potential.
By contrast, in a [001] QW we find that the non-conservation of
$S_z$ causes corrections to the side-jump  effect, as well as the
appearance of intrinsic contributions
 to the SHC from the nonlinear Dresselhaus term.

Thus, the spin Hall effect in a  [110] GaAs QW offers a hitherto
unexplored opportunity to measure ``pure" skew-scattering and
side-jump contributions. And since the skew-scattering spin-Hall
conductivity increases with increasing mobility while the
side-jump conductivity is not affected, we can propose a new
experiment, where the change in the sign of SHC with temperature
reveals the dominance of one or the other mechanisms.

{\it Model and results.} Our effective hamiltonian for
 the conduction band of a [110] QW includes Dresselhaus spin-orbit
couplings,  as well as SO corrections to the  electron-impurity and the electron-electron  (e-e) interactions:
\begin{eqnarray}\label{Hamiltonian}
\hat H &=&\sum_{i=1}^N \left[\frac{\hat {p}_i^2}{2 m^*} +
V(\hat{\vec r}_i)\right]+ \frac{1}{2}V_{e-e} \nonumber\\ &+&
\frac{\alpha_1}{\hbar} \sum_{i=1}^N  \left( \hat {p}_{ix}
\nabla_{iy} \hat V_{Ti}-\hat {p}_{iy} \nabla_{ix} \hat V_{Ti}
\right) \hat{S}_{iz} \nonumber\\ &+&\sum_{i=1}^N
\left[\frac{\alpha_2}{\hbar}eE_0\hat{p}_{iy}\hat{S}_{iz}
+\frac{\alpha'_2}{\hbar}eE_0(-\hat{p}_{iy}\hat{p}_{ix}^2+\hat{p}_{iy}^3/2)
\hat{S}_{iz}\right]~.\nonumber\\
\end{eqnarray}
Here $\hat {\vec p}_i$, $\hat {\vec r}_i$ are the canonical
momentum and position operators of the $i$-th electron,
 $V_{e-e}=\sum_{j \neq i}\frac{e^2}{\epsilon_b|\hat {\vec
r}_i-\hat{\vec r}_j|}$, $\hat V_{Ti}\equiv V(\hat {\vec r}_i) +
V_{e-e}$ is the total potential acting on the i-th electron due to
random impurities ($V(\vec r)$) and e-e interactions in the $(110)$
plane, $\hat {S}_{iz}$ is the Pauli matrix operator, $x$, $y$, and
$z$ are cartesian component indices with $z$ along the $[110]$ axis,
$m^*$ is the conduction band mass, $e$ is the absolute value of the electron charge, $E_0$ is a ``built-in"
electric field originating from the
 crystal symmetry, and the quantities
$\alpha_1,\alpha_2,\alpha'_2$  are the strengths of SO coupling in
the semiconductor, whose values can be calculated from the matrix
elements of the momentum operator between different bands within the
$14\times 14$ band model \cite{Winkler2003}. Notice that $\hat
S_{iz}$ is a constant of the motion.  Although $\alpha_1$, $\alpha_2$
and $\alpha'_2$ are connected, we assign them different labels in
order to (artificially) turn off one or the other effect in our calculations. These calculations are done to first order in
the $\alpha$'s, which is correct if energies associated with the
$\alpha$'s are much smaller than the Fermi energy.  Notice that the SO
 coupling energy is not required to be much smaller than
$\hbar/\tau$.

We perturb the system with a uniform electric field of frequency
$\omega$ in the $x$ direction, which is described by a vector
potential
$\frac{e}{c} \vec A(t) = \frac{e}{i\omega}E \vec e_x e^{-i \omega t}+c.c.~.$
The perturbed Hamiltonian, to first order in $E$, is
obtained from the canonical replacement $\hat {\vec p} \to \hat
{\vec p} + \frac{e}{c} \vec A(t)$, which gives
\begin{eqnarray}\label{H.perturbed}
\hat H(t) = \hat H +\frac{e}{c}\sum_{i=1}^N \left( \frac{\hat
p_{ix}}{m^*}+2\frac{\alpha'_2}{\hbar}eE_0\hat{p}_{iy}\hat{p}_{ix}\hat{S}_{iz}+
\right.\nonumber\\ \left.\frac{\alpha_1}{\hbar} \nabla_{iy}
V_{Ti}\hat {S}_{iz}\right) A_x(t)~.
\end{eqnarray}
We want to
calculate the magnitude of the transverse $z$-spin current defined
as $\hat J^z_y \equiv \frac{\hbar}{2} \sum_{i=1}^N\frac{\hat
v_{iy}\hat S_{iz} + \hat S_{iz}\hat v_{iy}}{2}$, where $\hat
v_{iy}$ is the $y$-component of the velocity operator. To find the
correct expression for  $\hat {\vec v}_{i}$ we note that, as a
result of the transformation from the the original multi-band
Hamiltonian to the effective Hamiltonian~Eq.~(\ref{H.perturbed}),
the physical position operator no longer coincides with the
canonical position operator $\hat {\vec r}_i$, but is given by
\footnote{For an eight band model the formula for the position
operator was derived in Ref.[12]. We found that the same formula
is valid for fourteen band model. }
\begin{equation} \label{r.effective}
\hat {\vec r}_{phys,i} = \hat{\vec r}_i- \frac{\alpha_1}{\hbar}
\left(\hat {\vec p}_i+\frac{e}{c} \vec A\right) \times \hat {\vec
S}_i~.
\end{equation}
The velocity is the time derivative of $\hat {\vec r}_{phys}$  and
making use of $-\frac{1}{c}\dot{\vec A} = \vec E$  we find (to
first order in the $\alpha$'s and dropping a ``diamagnetic" term
with zero average)
\begin{eqnarray}\label{v.physical}
&&\hat v_{iy}=\frac{1}{m^*}\hat{p}_{iy}-\frac{\alpha_1}{\hbar}
\nabla_{ix} V_{Ti}\hat{S}_{iz}\nonumber\\
&&+\left(\frac{\alpha_2}{\hbar}eE_0+\frac{\alpha'_2}{\hbar}eE_0(-\hat{p}_{ix}^2+\frac{3\hat{p}_{iy}^2}{2})
\right)\hat{S}_{iz}\nonumber\\
&&-\frac{\alpha_1}{\hbar} \nabla_{ix}
V_{Ti}\hat{S}_{iz}-\frac{\alpha_1 e}{\hbar}E_x\hat{S}_{iz}
\end{eqnarray}
It is essential for what follows that a steady state be
established in response to the d.c. electric field.
 In the steady state regime the average force  exerted on
electrons of either spin by impurities and other electrons
$\langle - \vec\nabla V_{T} \rangle$ must be exactly balanced by
the average force exerted by the electric field, $-e \vec E$.  So
for the average spin current, the last two terms of
Eq.~(\ref{v.physical}) cancel out and we get:
\begin{eqnarray}\label{Jzspin.3}
J^z_y =\frac{\hbar}{2}\left[\frac{\alpha_1n e}{\hbar} E_x
+\frac{\alpha_2n e}{\hbar} E_0+\right.\nonumber\\ \left.
\sum_{i=1}^N\left(\frac{\hat p_{iy}}{m}\hat
S_{iz}+\frac{\alpha'_2}{\hbar}eE_0(-\hat{p}^2
_{ix}+\frac{3}{2}\hat{p}^2 _{iy})\right)\right]
\end{eqnarray}
where $n$ is the electron density.  The first term on the r.h.s.
of Eq.~(\ref{Jzspin.3}) produces one half of the side-jump
contribution to the SHC, while the second one is the intrinsic
term generated by the internal field E$_0$. The SJ and intrinsic
terms have the same form. However, the term $\alpha_2neE_0/\hbar$
does not depend on the external electric field and because
$\langle J^z_y\rangle =0$ in the ground-state, it is cancelled by
the ground-state average of the other terms.

Now we use the linear response theory to calculate the
average spin current $\langle J^z_y \rangle$ to first order in $E$.   So we write $ J^z_y(t) =
\sigma^{SH}_{yx}(\omega) E e^{-i\omega t}+c.c.~$, where the
spin-Hall conductivity,  $\sigma^{SH}_{yx}(\omega)$ to the first
order in the strength of SO coupling is given by the
Kubo formula:
\begin{eqnarray}\label{linear.response.1}
&&\sigma^{SH}_{yx}(\omega)=\nonumber \\
&&\frac{1}{2}\alpha_1ne+\frac{e\hbar}{2i\omega {\cal
A}}\left[\frac{\alpha_1}{\hbar} \langle\langle \sum_{i=1}^N
\frac{\hat p_{iy}\hat S_{iz}}{m^*};\sum_{i=1}^N\nabla_{iy} \hat
V_{Ti}\hat{S}_{iz}\rangle\rangle\right.\nonumber\\
&&+\left.\langle\langle \sum_{i=1}^N \frac{\hat p_{iy}\hat
S_{iz}}{m^*};\frac{\hat {P}_x}{m^*}+2\frac{\alpha'_2}{\hbar}
eE_0\sum_{i=1}^N\hat{p}_{iy}\hat{p}_{ix}\hat{S}_{iz}\rangle\rangle
\right.\nonumber\\ &&+\left. \frac{\alpha'_2}{\hbar}\langle\langle
eE_0\sum_{i=1}^N(-\hat{p}_{ix}^2+\frac{3}{2}\hat{p}_{iy}^2);
\frac{\hat{P}_{x}}{m^*}\rangle\rangle\right]
\end{eqnarray}
where $\hat {\vec P}$ is the total momentum operator of the system,
 ${\cal A}$ is the area of the 2DEG, and  $ \langle \langle \hat
A;\hat B \rangle \rangle_\omega \equiv - \frac{i}{\hbar}
\int_0^\infty \langle [\hat A(t), \hat B(0) ] \rangle e^{-i \omega
t} dt $, where $\langle...\rangle$ is a short-hand for the Kubo linear response function.

We now show that  the second term on the r.h.s. of Eq.~(\ref{linear.response.1}) can be calculated exactly and, combined with the first term, gives the full
side-jump contribution to the SHC.  We first note that according to the Heisenberg equation of motion for the momentum operator, to zero order in $\alpha_1$ we can write $ \nabla_{iy} \hat V_{Ti}= -\frac{d}{dt}{\hat p}_{iy}$.  Furthermore $\hat S_{iz}$ is a strict constant of the motion, so we also have $ \nabla_{iy} \hat V_{Ti} \hat S_{iz}= -\frac{d}{dt}\left({\hat p}_{iy}\hat S_{iz}\right)$.   Making use of this identity   the  term in question can be rewritten as:
\begin{equation} -\frac{e\alpha_1}{2i\omega {\cal A}}
\langle\langle \sum_{i=1}^N \frac{\hat p_{iy}\hat
S_{iz}}{m^*};\sum_{i=1}^N\frac{d}{dt}\left({\hat p}_{iy}\hat S_{iz}\right)\rangle\rangle = -\frac{\alpha_1e}{2m^*}
\langle\langle\hat P_y;\hat P_y\rangle\rangle~,
\label{terminquestion}
\end{equation}
where in the last step we have used the well-known property of linear response functions,
\begin{equation}\label{correlator}
\langle \langle \hat A;\hat B \rangle \rangle_\omega
=-\frac{\langle \langle d\hat A/dt;\hat B \rangle
\rangle_\omega}{i \omega} +\frac{\langle [\hat A,\hat B]
\rangle}{\omega}~,
\end{equation}
and the fact that $\hat S_{iz}^2=1$.

The value of the $\langle\langle\hat P_y;\hat P_y\rangle\rangle$ response function in the $\omega \to 0$ limit is easily
obtained from the condition that the ordinary d.c. conductivity
\begin{equation}\label{sigmaxx}
\sigma_{yy}(\omega)=\frac{-1}{i{\cal A}\omega}
\left[\langle\langle -e\frac{\hat P_y}{m^*};-e\frac{\hat
P_y}{m^*}\rangle\rangle_\omega + \frac{Ne^2}
{m^*}\right]~
\end{equation}
is finite for $\omega \to 0$.\footnote{It is in this step
that we assume the system to be metallic.} This implies that the
quantity in the square brackets of Eq.~(\ref{sigmaxx}) vanishes in
the limit $\omega \to 0$, i.e.\footnote{Since, by translational
invariance, $\langle \langle\hat P_y;\hat
P_y\rangle\rangle_\omega=0$ in a perfectly clean system at any
finite frequency, such a cancellation can only occur in the
presence of disorder}\footnote{This result is exact in the absence
of spin-orbit couplings. Spin-orbit effects introduce negligible
correction of the order $E_{so}/E_F$, where $E_{so}$ is the
characteristic energy of spin-orbit interaction and $E_F$ is the
Fermi energy.}:
\begin{equation}\label{PxPx}
\lim_{\omega \to 0}\langle\langle \hat P_y;\hat P_y\rangle\rangle
= -Nm^* ~.
\end{equation}
Furthermore, it is easy to see that the the terms proportional to
$\alpha_2'$ in the last two lines of Eq.~(\ref{linear.response.1})
vanish for $\alpha_1=0$  (this is a consequence of the fact that
for $\alpha_1=0$ these terms are odd functions of $p_{ix}$ while
the Hamiltonian is  an even function of $p_{ix}$):  hence they are
at least of order $\alpha_2'\alpha_1$ and can be safely neglected.
 Then Eq.~(\ref{linear.response.1}), in the d.c. limit, simplifies to:
\begin{eqnarray}\label{linear.response.2}
\sigma^{SH}_{yx}=\alpha_1ne +\lim_{\omega \to 0}
\frac{e\hbar}{2i\omega {\cal A}}
\langle\langle \sum_{i=1}^N
\frac{\hat p_{iy}\hat S_{iz}}{m^*};\frac{\hat
{P}_x}{m^*}\rangle\rangle
%
~\end{eqnarray}
The first and the second term on the r.h.s. of this equation are naturally identified as the as
 the side-jump and the skew-scattering contributions to the SHC.
  Indeed the second term on the r.h.s. of  Eq.~(\ref{linear.response.2})
  is simply the response of the $y$-component of the canonical spin current to an electric
   field that couples to the $x$-component of the canonical particle current.
    This response includes neither the anomalous velocity nor the anomalous coupling to the electric field,
    and furthermore it vanishes for $\alpha_1=0$  (again, because of odd parity with respect to $p_{ix}$),
    hence it cannot be sustained by the band structure  alone.

Thus, we conclude that the total SHC for a [110] QW\cite{Sih05} is the sum of a
universal side-jump  contribution and a skew-scattering
contribution due to the impurities:
\begin{equation}\label{final.result0}
\sigma^{SH}_{yx}=\sigma^{sj}_{yx}+\sigma^{ss}_{yx}~,
\end{equation}
where
\begin{eqnarray}\label{final.result2}
\sigma^{ss}_{yx}&=& \lim_{\omega \to 0} \frac{e\hbar}{2i\omega
{\cal A}}\langle\langle \sum_{i=1}^N \frac{\hat p_{iy}\hat
S_{iz}}{m^*};\frac{\hat {P}_x}{m^*}\rangle\rangle~
\end{eqnarray}
and $\sigma^{sj}_{yx}=  \alpha_1ne$.

Notice that we have {\it not} assumed that either disorder or the
Coulomb interaction is weak: thus we have proved that the side
jump SHC in a [110] QW is independent of disorder and the Coulomb
potential  to all orders in the strength of these interactions.
The proof, however, depends on the conservation of $\hat S_z$,
which is a special feature of the  [110] QW.    When $\hat S_z$ is
not conserved, then Eq.~(\ref{terminquestion}) is no longer  true.
The spin precessional frequency, $\Omega_p$, while proportional to
$\alpha$,   will not be small in comparison to $1/\tau$ in the
clean limit.    As a result, terms that could be safely
disregarded when $\Omega_p$ was zero (conserved spin), become very
large when $\Omega_p \tau \gg 1$ and may even diverge when
$1/\tau$ tends to zero before $\Omega_p$.  This is the basic
mechanism through which the side-jump contribution can be modified
by spin precession and intrinsic contributions to the SHC can also
appear.  Indeed, in the case of a [001] GaAs QW we find that the
left hand side of Eq.~(\ref{terminquestion}) vanishes in the clean
limit, thus reducing the side-jump contribution to
$\frac{1}{2}\alpha_1ne$.

\begin{figure}[thb]
\includegraphics[width=3.4in]{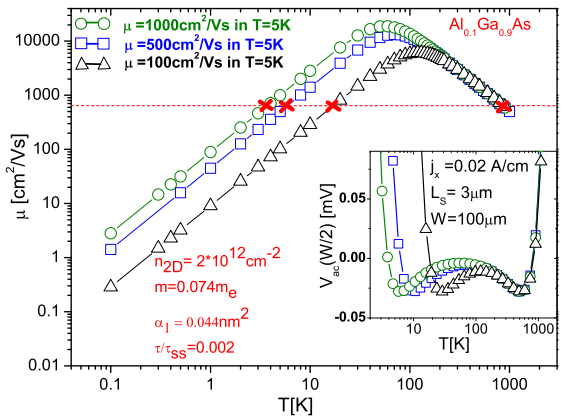}
\caption{(color online) Mobility, $\mu$, as a function of
temperature, $T$, for three different low-$T$ $\mu$'s. In inset, the
spin accumulation ($V_{ac}$) vs. $T$. The side jump contribution to
$V_{ac}$ dominates for low $T$. For increasing $T$, the lower
temperature red cross corresponds to the $T$ where the sign of
$V_{ac}$ starts to be controlled by skew scattering, the higher
temperature red cross to the place where side jump dominates again.}
\end{figure}
Let us now return to the [110] QW.  The skew-scattering contribution~(Eq.~\ref{final.result2})
is not easily obtained from perturbation theory, but we have recently shown, via the Boltzmann equation,
 that it is given by
 \begin{eqnarray}\label{final.result}
\sigma^{ss}_{yx}=  -\mu\frac{\hbar
n\tau/\tau_{ss}}{1+\gamma\tau}~,
\end{eqnarray}
where $\mu$ is the mobility,  $\tau_{ss}$ is the SS relaxation time
inversely proportional to $\alpha_1$, and $\gamma$ is the spin-drag
coefficient.\cite{Amico00,Weber05,Hankiewicz05}.   It is seen that e-e interactions are quite relevant here
 and reduce the SS term in
Eq.~(\ref{final.result}) by the factor $1+\gamma\tau$. Moreover,
the two contributions in Eq.~(\ref{final.result0})  have opposite
signs for an attractive impurity potential and the SS conductivity
increases with the mobility while the SJ is independent of it.
Similarly, the spin accumulation consists of two terms with
opposite signs and has the form \cite{Hankiewicz05}:
$V_{ac}(W/2)=-2L_sj_x\rho_D(\tau/\tau_{ss}-2e\alpha_1/\hbar\mu)\tanh(W/2L_s)$,
where $\rho_D$ is the Drude resistivity, $L_s$ is the spin
diffusion length, W is the width of sample and $j_x$ is the
current density \footnote{We omitted the spin Coulomb drag
correction which is small for the mobilities considered here.}. We
then propose a new experiment where the SJ/SS contributions can be
distinguished through the temperature dependence of
$\sigma^{SH}_{yx}$ or the spin accumulation potential $V_{ac}$.
Fig.1 presents the behavior of $\mu$ versus $T$ for experimentally
attainable samples. Notice that the values of parameters for the
theoretical curve designated by circles are exactly the same as
the values reported for the samples in the recent experiments
\cite{Sih05} on a [110] QW. The samples with lower mobilities can
be easily obtained by additional doping with Si inside the quantum
well \cite{Awschalom06}. Hence $\mu$ will grow as $T^{3/2}$ for
low $T$ as a result of scattering from ionized impurities and will
decrease as $T^{-3/2}$ for larger $T$ due to phonon scattering. It
is thus possible to observe two changes of sign of $V_{ac}$ moving
from low to high $T$s:  $\mu =1/(AT^{-3/2}+BT^{3/2})$, where $A$
is found from the low-$T$ mobility and $B$ is fixed by a room
temperature mobility of 0.3 m$^2$/Vs for AlGaAs. At low $T$ the
mobility is low and the SJ contribution to $V_{ac}$ dominates. The
first cross in the graph designates the point where the SS begins
to dominate, and the second cross, at still higher $T$, is the
point where the SJ retakes control of the sign of $V_{ac}$. Even
if the sign change is not detected, it is possible to tell whether
SS or SJ dominates by measuring whether $V_{ac}$ increases or
decreases as $\mu$ increases with changing $T$.

{\it Summary.} We have studied the spin-Hall effect for a [110] QW
taking into account the Dresselhaus SO coupling terms and the spin-orbit
interaction between electrons and impurities in the presence of
electron-electron interactions. We have shown that in the recent experiment
of Ref.~\cite{Sih05} the spin-Hall effect presents a clean competition between
side jump and skew scattering contributions (no intrinsic terms).
Furthermore we have
proposed a new experiment where the change in the sign of
spin-Hall accumulation with temperature reveals the dominance of
one or the other effect.  Finally, we have proved  that the
side-jump part of the spin-Hall conductivity is independent of disorder and
the Coulomb interaction at {\it all orders}, provided the spin current
is associated with a conserved component of the spin.

{\it Acknowledgements.} We thank Jairo Sinova for useful comments.
This work was supported by NSF Grant No. DMR-0313681 and DARPA/ONR
N00014-05-1-0913.


\end{document}